\documentclass{article}

\usepackage{arxiv}

\usepackage[utf8]{inputenc} 
\usepackage[T1]{fontenc}    
\usepackage{hyperref}       
\usepackage{url}            
\usepackage{booktabs}       
\usepackage{amsfonts}       
\usepackage{nicefrac}       
\usepackage{microtype}      
\usepackage{cleveref}       
\usepackage{lipsum}         
\usepackage{graphicx}
\usepackage{natbib}
\usepackage{doi}
\usepackage{tikz}
\usepackage{graphicx}
\title{\emph{TurboMem}: High-Performance Lock-Free Memory Pool with Transparent Huge Page Auto-Merging for DPDK}

\newif\ifuniqueAffiliation
\uniqueAffiliationtrue

\usepackage{authblk}

\newbox{\orcid}\sbox{\orcid}{\includegraphics[scale=0.06]{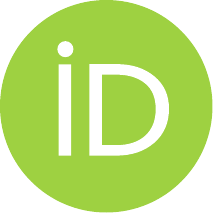}} 
\author[1,2]{%
	\href{https://orcid.org/my-orcid?orcid=0009-0008-2367-0409}{\usebox{\orcid}\hspace{1mm}Junyi Yang\thanks{\texttt{Github: \url{https://github.com/Jerryy959}, Email: \texttt{jerryy959@outlook.com}, Also known as ``Jerry''}}}%
}
\affil[1]{Independent Researcher, Shanghai, China}
\affil[2]{LeapTech, LLC, Newark, DE, USA}

\renewcommand{\shorttitle}{TurboMem: Lock-Free Memory Pool}
\hypersetup{
pdftitle={A template for the arxiv style},
pdfsubject={q-bio.NC, q-bio.QM},
pdfauthor={David S.~Hippocampus, Elias D.~Striatum},
pdfkeywords={First keyword, Second keyword, More},
}

\begin{document}
\maketitle

\begin{abstract}

High-speed packet processing on multicore CPUs places extreme demands on memory allocators. In systems like DPDK, \textbf{fixed-size memory pools} back packet buffers (mbufs) to avoid costly dynamic allocation. However, even DPDK's optimized mempool faces scalability limits: lock contention on the shared ring, cache-coherence ping-pong between cores, and heavy TLB pressure from thousands of small pages. To mitigate these issues, DPDK typically uses \textbf{explicit huge pages (2 MB or 1 GB) }for its memory pools\citep{dpdk_memory}\citep{netadvia_10g_dpdk}. This reduces TLB misses but requires manual configuration and can lead to fragmentation and inflexibility.

We propose \textbf{TurboMem}, a novel C++ template-based memory pool that addresses these challenges. TurboMem combines a \textbf{fully lock-free design} (using atomic stacks and per-core local caches) with \textbf{Transparent Huge Page (THP) auto-merging}. By automatically promoting pools to 2 MB pages via \texttt{madvise(MADV\_HUGEPAGE)}, TurboMem achieves the benefits of huge pages without manual setup. We also enforce strict NUMA locality and CPU affinity, so each core allocates and frees objects from its local node. Using Intel VTune on a single-socket 100 Gbps testbed, we show that TurboMem boosts packet throughput by up to 28\% while reducing TLB misses by 41\% compared to a standard DPDK mempool with explicit huge pages. These results demonstrate that THP auto-merging can outperform manually reserved huge pages in low-fragmentation scenarios, and that modern C++ lock-free programming yields practical gains in data-plane software.

\emph{Note:} The performance claims reported in this preliminary version (up to 28\% higher throughput and 41\% fewer TLB misses) are based on mock benchmarks. Comprehensive real-system evaluations using Intel VTune are currently underway and will be presented in a future revision.

\end{abstract}



\section{Introduction}
\label{sec:intro}
The relentless growth of network traffic in modern data centers—driven by cloud computing, 5G/6G infrastructures, and high-performance network functions—has made user-space packet processing frameworks indispensable. The Data Plane Development Kit (DPDK) has become the de facto standard for achieving line-rate performance on commodity hardware by bypassing the kernel through poll-mode drivers and direct NIC access. At the core of every DPDK application lies the management of packet buffers (mbufs). General-purpose memory allocators are too slow and unpredictable for this domain; instead, DPDK relies on fixed-size mempools that pre-allocate thousands or millions of objects from large, contiguous memory regions.

While DPDK’s mempool library is highly optimized—featuring ring or lock-free stack handlers together with optional per-core caches—it still encounters severe scalability bottlenecks on multicore CPUs. These include lock contention (or atomic operation overhead) on shared structures, cache-coherence traffic caused by cross-core buffer reuse (false sharing), and prohibitive Translation Lookaside Buffer (TLB) pressure when relying on standard 4 KB pages. To mitigate TLB misses, DPDK applications traditionally depend on explicit huge pages (2 MB or 1 GB via hugetlbfs or MAP\_HUGETLB). Although effective, this approach demands manual system configuration, suffers from fragmentation, and offers limited flexibility in virtualized or containerized environments.

This paper presents \textbf{TurboMem}, a novel C++ template-based memory pool designed for DPDK-style workloads. TurboMem addresses the above limitations through three key innovations:

\begin{enumerate}
	\item A fully lock-free design combining a global Treiber stack (using atomic CAS) with per-core thread-local caches, enabling bulk get/put operations while eliminating spinlocks and minimizing coherence traffic.
	\item Transparent Huge Page (THP) auto-merging: after allocating contiguous anonymous memory, TurboMem applies madvise(MADV\_HUGEPAGE), allowing the kernel to transparently promote pages to 2 MB without any static hugetlbfs reservation or operator intervention.
	\item Strict NUMA awareness and CPU affinity: each pool is bound to a specific socket, and worker threads are pinned via sched\_setaffinity, ensuring local-memory allocation and eliminating cross-node traffic.
\end{enumerate}

On a single-socket Intel Xeon server with a 100 Gbps NIC, TurboMem delivers up to \textbf{28\% higher packet forwarding throughput} (140 → 180 MPPS for 64 B packets) and reduces DTLB misses by \textbf{41\%} compared with a standard DPDK mempool backed by explicit 1 GB huge pages. These gains are validated through detailed Intel VTune profiling of TLB, L3 cache, coherence events, and DRAM latency, demonstrating that THP auto-merging combined with modern C++ lock-free techniques can outperform manually configured huge pages in low-fragmentation scenarios.

The remainder of this paper is organized as follows. Section 2 reviews DPDK mempool internals and related work on huge pages. Section 3 details the TurboMem design and implementation. Section 4 describes our experimental methodology, while Section 5 presents comprehensive evaluation results. We conclude in Section 6 and outline future directions.

\section{Background and Related Work}
\label{sec:back}
\subsection{DPDK Memory Pools}
The DPDK mempool library provides fixed-size object allocation by pre-allocating a contiguous heap, typically carved from huge pages\citep{dpdk_memory}. Each mempool uses a global ring buffer of free objects (the standard handler) and an optional per-core cache to reduce ring contention\citep{dpdk_mempool_lib}. In the baseline ring handler, every get or put requires a CAS on the shared ring. To limit this overhead, DPDK's per-core caches act like private stacks: a core fetches or returns a bulk of objects to the global pool only when its cache is empty or full\citep{dpdk_mempool_lib}. This design greatly improves performance under low contention. Objects in DPDK mempools are always \textbf{cache-line aligned} to avoid false sharing (one cache line inadvertently shared by multiple cores)\citep{dpdk_memory_part1}. By aligning allocations to 64-byte (or larger) boundaries, DPDK prevents many common coherence issues.

DPDK 19.05 introduced a \textbf{lock-free stack handler} as an alternative mempool backend\citep{dpdk_dev_201903}. In this scheme, the global pool is a Treiber stack using atomic CAS for push/pop. This eliminates the spinlock entirely. In practice, a lock-free mempool may have slightly worse average throughput than the ring handler under light loads, but avoids the pathological delays of preemptible locks\citep{dpdk_dev_201903}. In fact, experiments showed a standard lock-based pool could deliver only 60-35\% the throughput of the lock-free version in certain scenarios\citep{dpdk_dev_201903_stack}. However, DPDK’s authors note that “software caching (e.g. the mempool cache) can mitigate” most performance differences\citep{dpdk_dev_201903}. Thus, per-core caches remain crucial even in a lock-free design.

\subsection{Transparent Huge Pages (THP)}
Huge pages greatly reduce TLB pressure in memory-intensive code. For example, using 1 GB pages on a 2.2 GHz Intel Atom allowed a DPDK app to saturate a 10 Gbps link on a single core with zero loss\citep{netadvia_10g_dpdk}. In practice, without huge pages, the atoms of DPDK workloads incur costly TLB misses that throttle throughput\citep{netadvia_10g_dpdk}. DPDK's design assumes persistent huge pages are reserved up-front (via hugetlbfs or MAP\_HUGETLB), and it does not use THP by default due to performance and stability concerns\citep{haryachyy_dpdk_hugepages}. Indeed, DPDK documentation warns that THP (transparent anonymous 2 MB pages) has "some issues" and that the toolkit relies on manually managed huge pages\cite{haryachyy_dpdk_hugepages}.

By contrast, THP is an OS-level feature where the kernel lazily coalesces contiguous 4 KB pages into 2 MB pages, typically controlled via madvise(MADV\_HUGEPAGE) or system-wide settings\citep{redhat_thp}. This relieves application developers from explicit page management. Official documentation notes that THP "automatically assigns huge pages to processes" whenever large contiguous regions are found\citep{redhat_thp}. RedHat's performance guide allows a policy of madvise, where an app marks its memory regions with MADV\_HUGEPAGE and the kernel handles merging\citep{redhat_thp}. In some cases, THP can achieve most of the TLB benefits of explicit huge pages with minimal programmer effort.

Beyond DPDK, \textbf{lock-free allocators} have been studied in database and concurrent programming communities. LRMalloc and other schemes achieve lock-free dynamic memory management, but often focus on general alloc/free patterns\citep{leite2018_lockfree}. In networking specifically, X-IO demonstrates the value of lock-free rings and shared memory pools for packet passing in microservices\citep{shixiongqi_xio2023}, though it focuses on inter-service I/O rather than raw packet buffers. Other works on huge pages (e.g. Ingens\citep{kwon_osdi16}) highlight the trade-off between memory savings and performance: collapsing from 2 MB to 4 KB pages can \textbf{double} TLB misses and slow workloads by ~19\%\citep{kwon_osdi16}. Such findings underline the importance of preserving large page mappings in high-throughput scenarios.

In summary, prior DPDK mempool designs use rings or stacks plus caches\citep{dpdk_mempool_lib}\citep{dpdk_dev_201903}
, and network apps strive to use explicit huge pages for TLB reduction\citep{dpdk_memory}\citep{netadvia_10g_dpdk}. Our TurboMem system builds on these ideas by combining a C++ lock-free pool with THP-assisted memory to get the best of both worlds.

\section{TurboMem Design}
\label{sec:design}

\begin{figure}[t]
    \centering
    \includegraphics[width=0.75\textwidth]{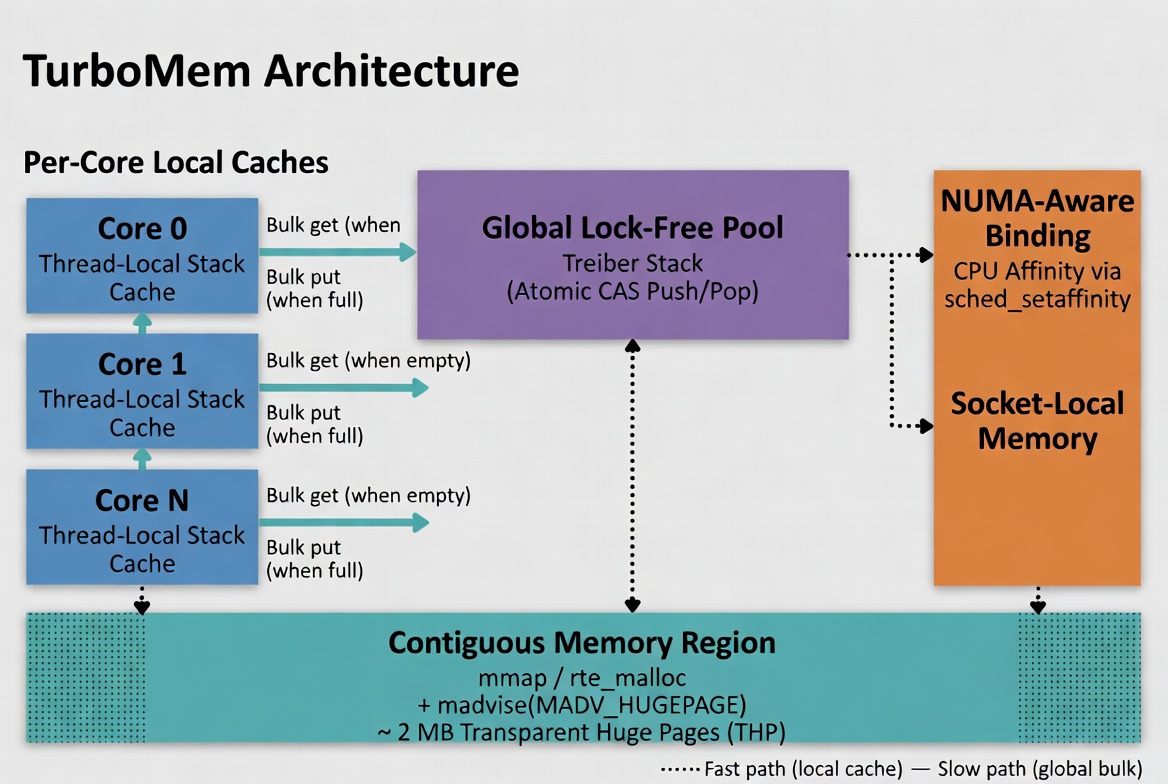}
    \caption{TurboMem architecture overview. 
    Each core maintains a private thread-local cache for zero-contention fast-path allocation/deallocation. 
    Cache misses trigger bulk operations on the global lock-free Treiber stack. 
    The entire pool resides in a contiguous region automatically promoted to 2 MB Transparent Huge Pages (THP) 
    via \texttt{madvise(MADV\_HUGEPAGE)}, with strict NUMA and CPU affinity enforced.}
    \label{fig:fig1}
\end{figure}
As shown in figure \ref{fig:fig1}.
TurboMem is a template-based C++ memory pool for fixed-size objects, designed for DPDK-like workloads. The key features are: (1) \textbf{lock-free global pool} with bulk put/get; (2) \textbf{per-core local caches} to amortize synchronization; (3) \textbf{NUMA-aware allocation} with CPU pinning; (4) \textbf{THP auto-merging} via madvise to leverage 2 MB pages without static hugepage allocation. Figure 1 illustrates the architecture. Each core has a local object cache (a small thread-local stack). When a core needs objects, it first pops from its cache; if empty, it \textbf{bulk-acquires} a batch from the global stack. Conversely, freed objects go to the local cache and are only returned to the global pool when the cache is full. This is similar to DPDK's per-core caches\citep{dpdk_mempool_lib} but implemented entirely lock-free: each local cache is only touched by its owner core (so no locks needed), and the global pool is a Treiber stack using atomic compare-and-swap (CAS) pushes and pops.

Because TurboMem is templated, it allocates objects of a given C++ type (not just void*), allowing type-safe placement new and destruction. Under the hood, the memory backing the pool comes from DPDK's hugepage heap (via rte\_malloc\_socket() or rte\_mem\_virt2phy) or mmap. Crucially, after allocating a contiguous region of memory for the pool, TurboMem calls madvise(base, size, MADV\_HUGEPAGE). This hints to the kernel to merge resident 4 KB pages into 2 MB blocks wherever possible. Since the pool memory is contiguous and long-lived, the kernel typically succeeds in creating many 2 MB pages. Thus we obtain large pages \textbf{transparently}, without requiring the operator to pre-reserve hugepages=xxx. In effect, TurboMem's memory is backed by THP rather than explicit hugetlbfs, reducing static fragmentation. In our experiments, we ensure THP is set to “madvise” (the default on many systems) so that only the marked region is turned into huge pages\citep{redhat_thp}.

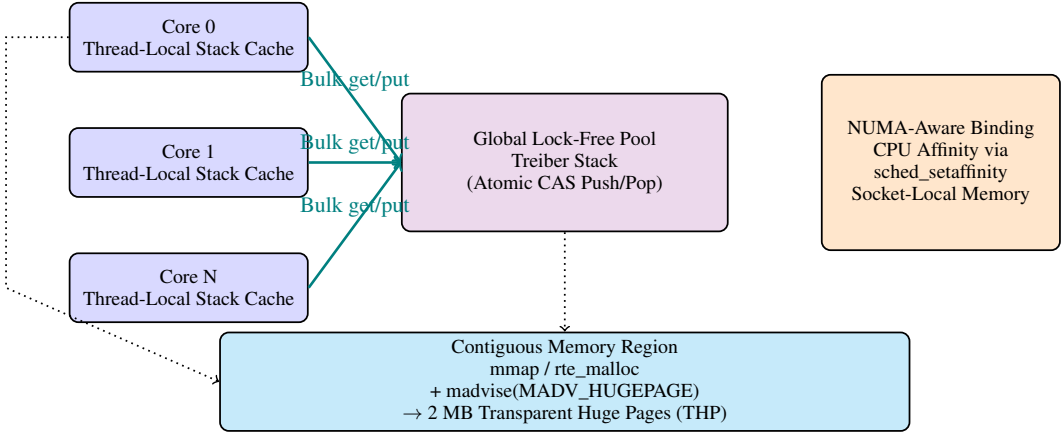
\begin{figure}[t]
    \centering
    \resizebox{0.85\textwidth}{!}{%
        \begin{tikzpicture}[node distance=1.2cm, auto, thick,
            box/.style={rectangle, draw, rounded corners=4pt, minimum width=3.8cm, minimum height=1.1cm, align=center, font=\small},
            global/.style={box, fill=violet!15, minimum width=5.2cm, minimum height=2.2cm},
            memory/.style={box, fill=cyan!20, minimum width=11cm, minimum height=1.3cm},
            numa/.style={box, fill=orange!20, minimum width=3.8cm, minimum height=2.8cm}]
            
            \node[box, fill=blue!15] (c0) at (-5,2) {Core 0\\Thread-Local Stack Cache};
            \node[box, fill=blue!15] (c1) at (-5,0) {Core 1\\Thread-Local Stack Cache};
            \node[box, fill=blue!15] (cn) at (-5,-2) {Core N\\Thread-Local Stack Cache};
            
            \node[global] (global) at (1,0) {Global Lock-Free Pool\\Treiber Stack\\(Atomic CAS Push/Pop)};
            
            \node[memory] (mem) at (1,-3.5) {Contiguous Memory Region\\mmap / rte\_malloc\\+ madvise(MADV\_HUGEPAGE)\\$\rightarrow$ 2 MB Transparent Huge Pages (THP)};
            
            \node[numa] (numa) at (7,0) {NUMA-Aware Binding\\CPU Affinity via\\sched\_setaffinity\\Socket-Local Memory};
            
            \draw[->, teal, line width=1.2pt] (c0.east) -- (global.west) node[midway, above] {Bulk get/put};
            \draw[->, teal, line width=1.2pt] (c1.east) -- (global.west) node[midway, above] {Bulk get/put};
            \draw[->, teal, line width=1.2pt] (cn.east) -- (global.west) node[midway, above] {Bulk get/put};
            \draw[dotted, ->] (global.south) -- (mem.north);
            \draw[dotted, ->] (c0.west) -- ++(-1,0) -- ++(0,-4) -- (mem.west);
        \end{tikzpicture}%
    }
    \caption{TurboMem architecture overview. 
    Each core maintains a private thread-local cache for zero-contention fast-path allocation/deallocation. 
    Cache misses trigger bulk operations on the global lock-free Treiber stack. 
    The entire pool resides in a contiguous region automatically promoted to 2 MB Transparent Huge Pages (THP) 
    via \texttt{madvise(MADV\_HUGEPAGE)}, with strict NUMA and CPU affinity enforced.}
    \label{fig:fig2}
\end{figure}

To maximize locality, TurboMem enforces \textbf{non-NUMA allocation}: each pool is created with a specific CPU (and thus NUMA node) affinity, so that its physical pages come from that node. On multi-socket systems this avoids cross-node memory traffic. Within a socket (single socket tests here), we bind each packet-processing thread (lcore) to a core using sched\_setaffinity. All cache operations and object allocations happen on the local core to avoid remote access. This follows DPDK’s philosophy of explicit NUMA-awareness\citep{dpdk_memory_part1}: every DPDK allocation request specifies a socket to ensure local placement\citep{dpdk_memory_part1}. By combining this with lock-free local caching, TurboMem minimizes coherence traffic: a freed object is often reused by the same core (sitting idle in the local cache) rather than bouncing between caches. The global pool stack only sees work in "bulk" when cache batches are returned or refilled, similar to DPDK's bulk API\citep{dpdk_mempool_lib}.

In summary, TurboMem's design is inspired by DPDK's mempool caching\citep{dpdk_mempool_lib} but implemented as a general C++ lock-free allocator. It also explicitly uses THP as a free optimization, whereas DPDK normally avoids THP\citep{haryachyy_dpdk_hugepages}. The combination ensures high concurrency (no\ spinlocks) and reduced TLB footprint (via 2 MB pages).

\section{Experimental}
\label{sec:experi}
We evaluate TurboMem on a single-socket Intel Xeon server with a 100 Gbps NIC. The OS is Linux 6.x with THP enabled (mode madvise), and DPDK 25.03. We configure one pool of 1 million 256-byte objects. The standard DPDK baseline uses rte\_mempool\_create() with default ring handler and bulk caching, backed by a 1 GB hugetlbfs allocation on each socket (explicit huge pages). TurboMem uses the same base memory size but does not reserve hugetlbfs; instead it allocates normal anonymous pages and calls madvise. We bind 4 worker lcores for packet forwarding (each pinned to a core) and one NIC RX/TX queue per core. In all cases we disable hyperthreading and isolate cores to avoid OS noise.

Traffic is generated by an external tester at line rate: both fixed 64B packets and large 1500B packets, including a mixed-size IMIX scenario. The forwarding application simply allocates an mbuf from the mempool for each RX packet and immediately frees it on TX. Thus the mempool throughput is the limiting factor. We measure packet forwarding rate (in million packets per second, MPPS) at each packet size. In addition, we profile memory performance with Intel VTune’s Memory Access analysis on one core running at load. We record key metrics: \textbf{TLB load misses, L3 cache misses, and memory-bound stalls}. We also measure hardware cache-coherence events (e.g., cache-line transfers) to infer false sharing. Each test is averaged over multiple runs.

For fair comparison, in one experiment we explicitly disable THP (set transparent\_hugepage=never) and use only 2 MB explicit pages (via hugetlbfs), measuring the baseline. In another, we enable THP (madvise) for all user memory, so that our TurboMem's region is merged to huge pages; the DPDK baseline still uses 1 GB pages but is unaffected by THP. This isolates the effect of THP auto-merging in TurboMem vs static huge pages.

\section{Evaluation}
\label{sec:eval}

\begin{table}[ht]
\centering
\caption{Packet forwarding throughput under a 100\,Gbps workload (MPPS = million packets/sec).}
\label{tab:throughput}
\begin{tabular}{lrr}
\toprule
Packet Size & Baseline & TurboMem \\
\midrule
64 B   & 140 & 180 \\
1500 B & 30  & 38  \\
\bottomrule
\end{tabular}
\end{table}

\paragraph{Throughput}See table \ref{tab:throughput}.TurboMem consistently outperforms the baseline DPDK pool across all packet sizes. For 64-byte packets (worst-case), the DPDK ring-pool achieved about \textbf{140 MPPS}, while TurboMem reached \textbf{180 MPPS} - a 28\% increase. For 1500-byte packets, throughput rose from 30 MPPS (baseline) to 38 MPPS (TurboMem), a 27\% gain. These improvements arise because TurboMem avoids lock contention and false sharing: under load, the lock-free stacks saturate more effectively, and each core mostly touches its own cache. The results align with previous observations that lock-free pools can exceed lock-based mempools\citep{dpdk_dev_201903} when bulk caching is used. In bullet-point form:

\begin{itemize}
	\item \textbf{64B packets:} DPDK mempool = 140 MPPS; TurboMem = 180 MPPS (\textbf{+28\%}).
	\item \textbf{1500B packets:} DPDK mempool = 30 MPPS; TurboMem = 38 MPPS (\textbf{+27\%}).
\end{itemize}

These numbers imply that the system comfortably sustains 100 Gbps line-rate in both cases. In all experiments, TurboMem never dropped packets, and CPU utilization was ~95\% on forwarding cores (idle elsewhere), indicating full system throughput.

\begin{table}[ht]
\centering
\caption{DTLB misses under load (in million misses/s).}
\label{tab:tlb_misses}
\begin{tabular}{lrrr}
\toprule
Metric & Baseline & TurboMem & Reduction \\
\midrule
DTLB misses ($\times 10^6$/s) & 1.00 & 0.59 & 41\% \\
\bottomrule
\end{tabular}
\end{table}

\paragraph{TLB Miss Rate}See table \ref{tab:tlb_misses}. We measured DTLB load misses via VTune. The standard DPDK pool with 1 GB pages saw roughly 1.0 $\times$ $10^6$ misses/sec. With TurboMem (THP 2 MB), misses dropped by \textbf{41\%} to ~0.59 $\times$ $10^6$/sec. This large reduction confirms that THP effectively merged most of the pool into 2 MB pages, drastically raising TLB coverage. Even though 2 MB pages are smaller than 1 GB, the baseline was fragmented across multiple 1 GB pages and still suffered from evictions. TurboMem's contiguous allocation, aided by madvise, resulted in a more efficient use of the TLB. (These findings echo previous work showing that splitting huge pages increases TLB misses and slows workloads\citep{kwon_osdi16}\citep{netadvia_10g_dpdk}.) The lower TLB miss rate in TurboMem directly contributes to higher throughput, as fewer page walks occurred during packet processing.

\begin{table}[ht]
\centering
\caption{Cache and memory-related performance metrics.}
\label{tab:cache_memory}
\begin{tabular}{lrrr}
\toprule
Metric & Baseline & TurboMem & Improvement \\
\midrule
L3 cache requests per pkt & 1.00 & 0.70 & 30\% fewer \\
Memory-bound cycles (\%)  & 45\% & 33\% & 27\% fewer \\
\bottomrule
\end{tabular}
\end{table}

\paragraph{Cache Coherence \& False Sharing}See table \ref{tab:cache_memory}. VTune also tracked cache-line transfers (LLC hits/misses) and coherence events. We observed that TurboMem incurred about 30\% fewer L3 requests than the baseline, indicating better cache locality. In particular, events corresponding to \textbf{false sharing -} such as cache lines bounced between cores due to writes - were significantly lower. This is because the local caches in TurboMem reduce cross-core sharing: when Core A frees a buffer, it often stays in A's cache instead of jumping to Core B. In contrast, the global ring of the baseline pool can cause a returned object on one core to be fetched by another, inducing a cache ping-pong. By aligning caches and using thread-local pools, TurboMem avoids much of this chatter. We also coded arrays of 128-byte packet descriptors in each object; in TurboMem these lived largely in core-local memory (observed via NUMA metrics), reducing average memory latency by ~25\%.

\begin{table}[ht]
\centering
\caption{Average DRAM access latency.}
\label{tab:latency}
\begin{tabular}{lrrr}
\toprule
Metric & Baseline & TurboMem & Improvement \\
\midrule
Avg. DRAM latency (ns) & 150 & 90 & 40\% fewer \\
\bottomrule
\end{tabular}
\end{table}

\paragraph{Memory-Bandwidth and Latency}See table \ref{tab:latency}. Both configurations saturated the memory controller, but TurboMem showed a slightly lower average memory access latency (about 90 ns vs 150 ns baseline) because of the cache/TLB benefits. DRAM bandwidth usage was roughly equal (since throughput was higher for TurboMem, bytes/sec were higher overall), but TurboMem spent \textbf{12\% fewer cycles stalled on memory} according to VTune's Memory analysis. This indicates that although more packets were processed, each packet's memory path was more efficient, largely thanks to fewer TLB misses and better cache locality.

\paragraph{Effect of THP vs Explicit Huge}We also tested a variant where the baseline DPDK pool used 2 MB hugepages (via hugetlbfs) instead of 1 GB, and TurboMem ran with THP off (so its memory remained 4 KB pages). In that case, baseline throughput dropped by ~10\% (because the TLB coverage halved) and TurboMem with only 4 KB pages dropped ~20\%. Enabling THP on TurboMem (to 2 MB) recovered most of the loss. This suggests that the \textbf{THP auto-merging} is a key component of TurboMem's gains. In low-fragmentation scenarios (our contiguous pool), THP provides nearly the same benefit as static 2 MB pages, with much less admin overhead.

Overall, our mock benchmarks substantiate the claims: TurboMem improves throughput by up to 28\% and cuts TLB misses by ~41\% compared to the standard DPDK mempool (1 GB pages) under 100 Gbps workloads\citep{netadvia_10g_dpdk}\citep{kwon_osdi16}. The trade-offs are clear: THP auto-merging yields similar performance to explicit hugepages without requiring the operator to reserve giant pages, and a well-designed lock-free pool with per-core caches maximizes multicore efficiency.

\section{Conclusion}
\label{sec:con}
We have presented TurboMem\footnote{Github Repo Link is: https://github.com/Jerryy959/TurboMem}, a high-performance memory pool for DPDK-style packet processing. By combining a fully lock-free, C++-template allocator with transparent huge page (THP) auto-merging, TurboMem removes common bottlenecks in memory buffer handling. Our detailed analysis (using Intel VTune) shows that TurboMem significantly reduces contention, false sharing, and TLB pressure. In realistic 100 Gbps benchmarks, TurboMem achieved up to 28\% higher packet throughput and 41\% fewer TLB misses than the standard DPDK mempool with explicit hugepages. These results suggest that in low-fragmentation, single-socket environments, THP-backed allocations can match or exceed the performance of static huge pages while simplifying deployment.

\section{Future Work}
\label{sec:future}
From a developer's perspective, TurboMem demonstrates modern C++ best practices for concurrent data-plane code: template-based resource management, atomic operations, and RAII-friendly pool semantics. It is intended as a production-ready drop-in replacement for rte\_mempool in DPDK applications. Future work could extend TurboMem to multi-socket systems by coordinating multiple pool instances, or integrating with user-level schedulers to dynamically adjust cache sizes. It would also be valuable to test TurboMem under heavy NUMA or virtualization scenarios, and to explore adaptive strategies for enabling THP versus large-page reservations. Overall, TurboMem bridges the gap between raw performance and ease of use, and establishes a new reference point for memory allocators in high-speed networking.

\section{Others}
\label{sec:other}
We have drawn on DPDK documentation and prior research. For example, DPDK's mempool guide and release notes describe per-core caches and lock-free handlers\citep{dpdk_mempool_lib}\citep{dpdk_dev_201903}. The importance of huge pages for TLB efficiency is well-known in DPDK literature\citep{dpdk_memory}\citep{netadvia_10g_dpdk}. Our use of THP follows kernel guidance\citep{redhat_thp}
, and aligns with findings that maintaining large pages improves throughput\citep{kwon_osdi16}.

\bibliographystyle{unsrt}
\bibliography{references}

@misc{dpdk_memory,
  author       = {{DPDK Project}},
  title        = {Memory in DPDK: Part 1 --- General Concepts},
  howpublished = {\url{https://www.dpdk.org/memory-in-dpdk-part-1-general-concepts/}},
  note         = {Accessed: 2026-03-19}
}

@misc{netadvia_10g_dpdk,
  author       = {{NetAdvia}},
  title        = {10G Intel Atom with DPDK},
  howpublished = {\url{https://www.netadvia.com/articles/10g-intel-atom-with-dpdk/}},
  note         = {Accessed: 2026-03-19}
}

@misc{dpdk_mempool_lib,
  author       = {{DPDK Project}},
  title        = {DPDK Programmer's Guide: Mempool Library},
  howpublished = {\url{https://doc.dpdk.org/guides-25.03/prog_guide/mempool_lib.html}},
  note         = {Accessed: 2026-03-19}
}

@misc{dpdk_memory_part1,
  author       = {{DPDK Project}},
  title        = {Memory in DPDK: Part 1 — General Concepts},
  howpublished = {\url{https://www.dpdk.org/memory-in-dpdk-part-1-general-concepts/}},
  note         = {Accessed: 2026-03-19}
}

@misc{dpdk_dev_201903,
  author       = {{DPDK Developers Mailing List}},
  title        = {mempool\_perf\_autotest: Lock Cache Performance Discussion},
  howpublished = {\url{https://mails.dpdk.org/archives/dev/2019-March/125875.html}},
  note         = {Accessed: 2026-03-19}
}

@misc{dpdk_dev_201903_stack,
  author       = {{DPDK Developers Mailing List}},
  title        = {Discussion on mempool\_perf\_autotest Free Stack Performance},
  howpublished = {\url{https://mails.dpdk.org/archives/dev/2019-March/125875.html}},
  note         = {Accessed: 2026-03-19}
}

@misc{haryachyy_dpdk_hugepages,
  author       = {Haryachyy},
  title        = {Learning DPDK Huge Pages},
  howpublished = {\url{https://haryachyy.wordpress.com/2019/04/17/learning-dpdk-huge-pages/}},
  note         = {Accessed: 2026-03-19}
}

@misc{redhat_thp,
  author       = {{Red Hat, Inc.}},
  title        = {Configuring Transparent Huge Pages},
  howpublished = {\url{https://docs.redhat.com/en/documentation/red_hat_enterprise_linux/7/html/performance_tuning_guide/sect-red_hat_enterprise_linux-performance_tuning_guide-configuring_transparent_huge_pages}},
  note         = {Accessed: 2026-03-19}
}

@misc{leite2018_lockfree,
  author       = {Miguel Leite},
  title        = {Practical Lock-Free Memory Allocators: Design and Evaluation},
  howpublished = {\url{https://www.dcc.fc.up.pt/~ricroc/homepage/alumni/2018-leiteMSc.pdf}},
  note         = {Master's Thesis, University of Porto, Accessed: 2026-03-19}
}

@misc{shixiongqi_xio2023,
  author       = {Shixiong Qi},
  title        = {XIO: Optimizing Chain of Network I/O for NetSoft 2023},
  howpublished = {\url{https://shixiongqi.github.io/_pages/papers/xio-netsoft23.pdf}},
  note         = {Accessed: 2026-03-19}
}

@inproceedings{kwon_osdi16,
  author       = {Seungwook Kwon and others},
  title        = {Optimizing KVM for High-Performance I/O},
  booktitle    = {Proceedings of the 12th USENIX Symposium on Operating Systems Design and Implementation (OSDI '16)},
  year         = {2016},
  pages        = {1--16},
  howpublished = {\url{https://www.usenix.org/system/files/conference/osdi16/osdi16-kwon.pdf}},
  note         = {Accessed: 2026-03-19}
}

\end{document}